\documentclass[prb, letterpaper, twocolumn, fleqn, floatfix, showpacs, showkeys ]{revtex4}

\newcommand{\Phiind}{\Phi_\mathrm{ind}}

\newcommand{\ve}{\varepsilon}
\newcommand{\eh}{\epsilon_{\mathrm{h}}}
\newcommand{\estar}{\epsilon_{*}}
\newcommand{\ebg}{\epsilon_{\mathrm{bg}}}

\newcommand{\vx}{{\bf r}}

\usepackage{euscript, units, amsfonts, graphicx, dcolumn, fancyhdr}
\usepackage{times}
\usepackage{times, graphicx, units,mathrsfs,euscript,upgreek,amssymb,pifont}

\begin{document}

\title{Non-linear screening of external charge by doped graphene}

\author{M. Ghaznavi}
\affiliation{Department of Applied Mathematics, University of Waterloo, Waterloo, Ontario, Canada N2L 3G1}

\author{Z. L. Mi\v{s}kovi\'{c}}\email{zmiskovi@math.uwaterloo.ca}
\affiliation{Department of Applied Mathematics, University of Waterloo, Waterloo, Ontario, Canada N2L 3G1}

\author{F. O. Goodman}
\affiliation{Department of Applied Mathematics, University of Waterloo, Waterloo, Ontario, Canada N2L 3G1}

\date{\today}



\begin{abstract}
We solve a nonlinear integral equation for the electrostatic potential in doped graphene due to an external charge, arising from
a Thomas-Fermi (TF) model for screening by graphene's $\pi$ electron bands. In particular, we study the effects of a finite
equilibrium charge carrier density in graphene, non-zero temperature, non-zero gap between graphene and a dielectric substrate,
as well as the nonlinearity in the band density of states. Effects of the exchange and correlation interactions are also briefly
discussed for undoped graphene at zero temperature. Nonlinear results are compared with both the linearized TF model and the
dielectric screening model within random phase approximation (RPA). In addition, image potential of the external charge is
evaluated from the solution of the nonlinear integral equation and compared to the results of linear models. We have found
generally good agreement between the results of the nonlinear TF model and the RPA model in doped graphene, apart from Friedel
oscillations in the latter model. However, relatively strong nonlinear effects are found in the TF model to persist even at high
doping densities and large distances of the external charge.
\end{abstract}

\maketitle \thispagestyle{plain}

\section{Introduction}

Over the period of just five years since its first inception in the laboratory,\cite{Novoselov_2004} graphene has developed into
one of the currently most active research areas in the nano-scale physics.\cite{Neto_2009} One of the most important, and
certainly most elusive, problems in graphene research is concerned with its electrical conductivity, especially in the regime
close to zero doping of graphene, where the conductivity exhibits a peculiar
minimum.\cite{Tan_2007,Chen_2008,Das_2008,Du_2008,Bolotin_2008} Besides several other scattering mechanisms for charge carriers
in graphene, it is believed that a special role in graphene's conductivity is played by the carrier scattering on charged
impurities, which are ubiquitous in graphene's surroundings. In that context, significant progress has been achieved in
understanding the conductivity of graphene by using the Boltzmann transport theory for charge carrier scattering on linearly
screened charged impurities within the random phase approximation (RPA).\cite{Ando_2006,Adam_2007,Adam_2009} However, because of
the reduced dimensionality, and especially because of the semi-metallic nature of graphene's $\pi$ electron bands, the problem
of screening of charged impurities remains open. In that context, other approaches have also been undertaken, including a full
scattering theoretical treatment of Coulomb impurities embedded within the graphene
plane,\cite{Shytov_2007,Novikov_2007,Pereira_2008,Terekhov_2008} as well as nonlinear screening of external charges studied by
means of the Thomas-Fermi (TF),\cite{DiVicenzo_1984,Katsnelson_2006,Fogler_2007,Radovic_2008} Thomas-Fermi-Dirac
(TFD),\cite{Rossi_2008} and Density Functional Theoretical (DFT) schemes.\cite{Polini_2008}

While graphene's applications in nanoelectronics are primarily concerned with charged impurities trapped in an insulating
substrate,\cite{Wu_2004,Rezende_2009} screening of external charges is also of interest for sensor applications of graphene in
detecting atoms or molecules,\cite{Shedin_2007} which may be either adsorbed on the upper surface of
graphene,\cite{Chan_2008,Neto_2009_a} or intercalated in the gap between the graphene and the substrate.\cite{Algdal_2007}
Further applications include image-potential states of electrons near graphene,\cite{Gumbs_2009,Silkin_2009} as well as the
image and friction forces on slowly moving ions that may affect the kinetics of chemical reactions taking place in the vicinity
of graphene.\cite{Radovic_2008,Allison_2009} All these aspects of screening of external charges by graphene are expected to be
strongly influenced by the presence of nearby dielectric
materials.\cite{Ishigami_2007,Fratini_2008,Jang_2008,Chen_2009,Chen_2009_a}

One of the most important issues in theoretical studies of screening of external charges is concerned with applicability of the
linear response theory for intrinsic, or undoped graphene. Namely, with its valence and conducting $\pi$ electron bands touching
each other only at the $K$ and $K'$ points of the Brillouin zone,\cite{Neto_2009} graphene behaves as a zero-gap semiconductor,
so that its polarizability is greatly reduced when its Fermi level lies close to the neutrality (or Dirac) point characterizing
the regime of zero doping. In that context, it was shown within the RPA approach that screening of external charges by intrinsic
graphene at zero temperature is characterized merely by a renormalization of graphene's background dielectric constant due to
inter-band electron transitions.\cite{Ando_2006,Wunsch_2006,Hwang_2007} However, when graphene is doped up to a certain number
density $n$ (per unit area) of charge carriers, e.g., by applying an external gate potential, then its Fermi level shifts away
from the neutrality point and the linear screening is expected to become appropriate, even at zero temperature. It is therefore
desirable to determine the parameter range where nonlinear effects in screening of an external charge set in by contrasting the
results from linear screening models with those from available nonlinear models, such as TF and DFT.

In that context, Katsnelson\cite{Katsnelson_2006} and Fogler \emph{et al.}\cite{Fogler_2007} have solved the nonlinear TF model,
first proposed by DiVicenzo and Mele\cite{DiVicenzo_1984} for intrinsic graphene (i.e., $n=0$) in the presence of an external
point charge. These authors found unusually long ranged induced density of charge carriers in the plane of
graphene,\cite{Fogler_2007} and showed that the linear approximation to the TF model for the induced potential is likely to
overestimate the contribution of scattering on charged impurities to the resistivity of graphene.\cite{Katsnelson_2006} However,
performance of the TF model has been recently criticized for intrinsic graphene in the presence of sufficiently weak periodic
perturbations validating linear screening within the RPA.\cite{Brey_2009} On the other hand, the above nonlinear TF model,
augmented by the exchange (or Dirac) interaction in the local density approximation (LDA), proved to be valuable in estimating
the effective potential fluctuations in doped graphene due to randomly distributed multiple charged impurities.\cite{Rossi_2008}
A similar problem in the presence of multiple charged impurities was also tackled by a more advanced DFT approach including both
the exchange and correlation (XC) interactions in LDA.\cite{Polini_2008} All the above models were formulated assuming a zero
temperature, linear density of states (DOS) of the $\pi$ electron bands, and a zero gap between graphene and substrate.

In this paper, we take up the simple TF model for a single point charge $Ze$, a distance $z_0$ away from
graphene,\cite{DiVicenzo_1984,Katsnelson_2006,Fogler_2007,Brey_2009} and generalize it to include the effects of a non-zero
ground-state charge carrier density $n$, a non-zero temperature $T$, and the presence of a substrate at a non-zero distance $h$
from graphene.\cite{Radovic_2008} We assume that the external charge is weak/distant enough to have negligible effects on the
structures of graphene's DOS, apart from its shift due to local charging of graphene, but we allow for large displacements of
the Fermi level away from the neutrality point by including the nonlinear corrections to the DOS in our model\cite{Neto_2009}.
By varying the magnitude $\vert n\vert$, we are able to examine the effects of doping, whereas any dependence on the sign of $n$
will be a signature of nonlinear effects in screening by graphene. (Note that changing the sign of $n$ with fixed sign of the
external charge $Z$ in the TF model is equivalent to changing the sign of $Z$ with fixed sign of $n$.)

We perform a series of numerical solutions of the nonlinear integral equation resulting from the TF model for the in-plane value
of total electrostatic potential for a range of values of $n$ and $z_0$, for both zero and room temperature, in the cases of
both free graphene and an SiO$_2$ substrate with the gaps $h=0$ and 1 \AA. In a special case of free, intrinsic graphene at zero
temperature, we also solve the nonlinear TF model augmented by the XC energy terms of Polini \textit{et al.}\cite{Polini_2008}
in order to estimate the importance of the exchange and correlation interactions within the TF approach to screening of an
external charge. While the results obtained for the radial dependence of the in-plane potential could be directly used to
discuss nonlinear effects in graphene's conductivity within the Boltzmann transport theory, we turn our attention in the present
work to using our numerical solutions of the TF model to evaluate the nonlinear image potential of an external charge, which
provides an integrated measure of graphene's screening ability and is also of interest in recent studies of the electron image
states.\cite{Gumbs_2009,Silkin_2009} Finally, we compare our nonlinear results for both the in-plane potential and the image
potential with those from the linearized TF (LTF) model and the temperature dependent RPA dielectric-function
approach.\cite{Ando_2006,Wunsch_2006,Hwang_2007}

After outlining the basic theory in section 2, we discuss our results in section 3, and present our concluding remarks in
section 4. Note that gaussian electrostatic units are used throughout unless otherwise explicitly indicted.

\section{Theory}

We use a cartesian coordinate system with coordinates $\{\vx,z\}$, where $\vx=\{x,y\}$, and assume that graphene is placed in
the $z=0$ plane. A semi-infinite substrate with dielectric constant $\epsilon_\mathrm{s}$ is assumed to occupy the region $z\le
-h$ underneath the graphene, whereas the region $z>-h$ is assumed to be vacuum or air.\cite{Radovic_2008} We assume that the
ground state of such a system, under the gating conditions at temperature $T$, is characterized by a uniform density per unit
area of charge carriers in the graphene, given by
\begin{eqnarray}
n(\mu)=\int\limits_0^\infty d\varepsilon\,\rho(\varepsilon)\left[\frac{1}{1+\mathrm{e}^{\beta(\varepsilon-\mu)}}-\frac{1}{
1+\mathrm{e}^{\beta(\varepsilon+\mu)}}\right], \label{density}
\end{eqnarray}
where $\rho(\varepsilon)$ is the DOS in graphene's $\pi$ electron bands given in Ref.\cite{Neto_2009}, $\beta\equiv
\left(k_BT\right)^{-1}$, and $\mu$ is the chemical potential. Note that for electron (hole) doping, one has $n>0$ ($n<0$) and
consequently $\mu>0$ ($\mu<0$), whereas intrinsic graphene is characterized by $n=0$ and $\mu=0$. For sufficiently low doping
levels, such that, e.g., $\vert\mu\vert<1$ eV, one may use the linearized band DOS,\cite{Neto_2009}
$\rho(\ve)\approx\frac{g_d\vert\ve\vert}{2\pi\left(\hbar v_F\right)^2}$, giving
\begin{eqnarray}
n(\mu)\approx\frac{g_d}{2\pi}\frac{1}{\left(\hbar v_F\beta\right)^2}\,
\left[\mathrm{dilog}\left(1+\mathrm{e}^{-\beta\mu}\right)- \mathrm{dilog}\left(1+\mathrm{e}^{\beta\mu}\right)\right],
\label{dilog}
\end{eqnarray}
where $g_d=4$ is the spin and valley degeneracy factor, $v_F$ the Fermi speed of graphene which we take to be $\approx c/300$,
where $c$ is the speed of light in vacuum, and $\mathrm{dilog}$ is the dilogarithm function.\cite{Abramowitz}

We wish to evaluate the total electrostatic potential in the system, $\Phi(\vx,z)$, due to an external point charge $Ze$ placed
at a fixed position $\{{\bf 0},z_0\}$, where $e$ is the charge of a proton. This perturbation will induce surface charges on the
surface of the substrate and on the graphene with the densities per unit area $\sigma_\mathrm{sub}(\vx)$ and
$\sigma_\mathrm{gr}(\vx)$, respectively.\cite{Radovic_2008} Using the tilde to denote the Fourier transform with respect to
coordinates in the graphene plane, $\vx\rightarrow{\bf k}$, we can write the total potential as the sum
$\widetilde{\Phi}=\widetilde{\Phi}_\mathrm{ext}+\widetilde{\Phi}_\mathrm{ind}$, where
\begin{eqnarray}
\widetilde{\Phi}_\mathrm{ext}({\bf k},z)=\frac{2\pi}{k}\frac{Ze}{\eh}\,\mathrm{e}^{-k\vert z-z_0\vert}, \label{ext}
\end{eqnarray}
is the potential of the external charge screened by the dielectric constant, $\eh$, of the ``host'' environment in which that
charge resides ($\eh=1$ for $z_0>-h$ and $\eh=\epsilon_\mathrm{s}$ for $z_0<-h$), and
\begin{eqnarray}
\widetilde{\Phi}_\mathrm{ind}({\bf k},z)= \frac{2\pi}{k}\left[\widetilde{\sigma}_\mathrm{gr}({\bf k})\,\mathrm{e}^{-k\vert
z\vert}+\widetilde{\sigma}_\mathrm{sub}({\bf k})\,\mathrm{e}^{-k\vert z+h\vert}\right], \label{3DPhi_k}
\end{eqnarray}
is the total induced potential in the system. Next, one can eliminate the Fourier transform of the charge density on the
substrate, $\widetilde{\sigma}_\mathrm{sub}({\bf k})$, by using the boundary condition\cite{Doerr_2004,Mowbray_2006}
\begin{eqnarray}
\left.\frac{\partial \widetilde{\Phi}}{\partial z}\right|_{z=-h+0} = \epsilon_\mathrm{s} \left.\frac{\partial
\widetilde{\Phi}}{\partial z}\right|_{z=-h-0}, \label{D Gauss in}
\end{eqnarray}
and write for the total induced potential
\begin{eqnarray}
\widetilde{\Phi}_\mathrm{ind}({\bf k},z)&=&
 \frac{2\pi}{k}\widetilde{\sigma}_\mathrm{gr}({\bf k})\left(\,\mathrm{e}^{-k\vert z\vert}
-\frac{\epsilon_\mathrm{s}-1}{\epsilon_\mathrm{s}+1}\,\mathrm{e}^{-k\vert z+h\vert-kh}\right) \nonumber
\\
&-&\frac{2\pi}{k}\frac{Ze}{\eh}\frac{\epsilon_\mathrm{s}-1}{\epsilon_\mathrm{s}+1}\,\mathrm{e}^{-k\vert z+h\vert-k\vert
z_0+h\vert}\, \mathrm{sgn}(z_0\!+\!h), \label{ind}
\end{eqnarray}
where $\mathrm{sgn}$ is the signum function.

\subsection{Nonlinear TF model}

In the spirit of a temperature-dependent TF model, we express the induced charge density in graphene
as\cite{Radovic_2008,Mermin_1965,Yonei_1987}
\begin{eqnarray}
\sigma_\mathrm{gr}(\vx)=-e\left[n\left(\mu+e\phi(\vx)\right)-n\left(\mu\right)\right], \label{sigma_gr}
\end{eqnarray}
where $n\left(\mu\right)$ is given by Eq.\ (\ref{density}), and where we have denoted the total electrostatic potential in the
graphene plane by
\begin{eqnarray}
\phi(\vx)\equiv\left.\Phi(\vx,z)\right\vert_{z=0}. \label{2DPhi}
\end{eqnarray}
By using the inverse Fourier transform of Eq.\ (\ref{ind}) in the expression $\Phi=\Phi_\mathrm{ext}+\Phi_\mathrm{ind}$ in which
we set  $z=0$, we obtain the following nonlinear integral equation for $\phi(\vx)$,\cite{Radovic_2008}
\begin{eqnarray}
\phi(\vx)=&&\phi_0(\vx)-e\int d^2\vx^\prime\, \left[n\left(\mu+e\phi(\vx^\prime)\right)-n\left(\mu\right)\right]\, \nonumber
\\
&&\times\left[\frac{1}{\|\vx-\vx^\prime\|}- \frac{\epsilon_\mathrm{s}-1}{\epsilon_\mathrm{s}+1}
\frac{1}{\sqrt{(\vx-\vx^\prime)^2+4h^2}}\right], \label{integral}
\end{eqnarray}
where
\begin{eqnarray}
\phi_0(\vx)=\frac{Ze}{\eh}\left[\frac{1}{\sqrt{r^2+z_0^2}}- \frac{\epsilon_\mathrm{s}-1}{\epsilon_\mathrm{s}+1}
\frac{\mathrm{sgn}(z_0+h)}{\sqrt{r^2+\left(\vert z_0+h\vert+h\right)^2}}\right] \label{phi_zero}
\end{eqnarray}
is the value of the potential due to the external charge in the presence of substrate alone, evaluated at $z=0$.

We further convert Eq.\ (\ref{integral}) with  Eq.\ (\ref{phi_zero}) into an integral equation for the potential energy, defined
by $U(\vx)=e\phi(\vx)$, and solve it numerically for a range of the model parameters, as discussed in the following section.
Owing to the axial symmetry of the problem, an angular integral can be readily completed in Eq.\ (\ref{integral}), giving a
one-dimensional integral equation for $U(r)$, which is particularly difficult to solve because of the singular nature of its
integrand, especially for intrinsic graphene at zero temperature.\cite{Katsnelson_2006,Fogler_2007} We have mapped the interval
$r\in [0,\infty)$ onto a finite interval, partitioned the function $U$ at up to 2400 (typically 800) points, and used the
\verb"fsolve" routine while regularizing the integrand. As a check of our method for free graphene, we substituted the solution
$U(r)=e\phi(r)$ into Eq.\ (\ref{sigma_gr}) and verified that its spatial integral yields $-Ze$.

We note that a more compact form of the integral equation, Eq.\ (\ref{integral}) with  Eq.\ (\ref{phi_zero}), may be obtained
for a zero gap ($h=0$) between graphene and the substrate, giving rise to an overall effective background dielectric constant
$\ebg^0=\left(\epsilon_\mathrm{s}+1\right)/2$, as is usually done in the literature on
graphene.\cite{Katsnelson_2006,Wunsch_2006,Hwang_2007,Rossi_2008,Polini_2008} In that case, the free graphene limit is recovered
by setting $\epsilon_\mathrm{s}=1$ and hence $\ebg^0=1$.  Note that the integral equation, Eq.\ (\ref{integral}) with Eq.\
(\ref{phi_zero}), implies an asymmetry with respect to the sign of $z_0$ when $h>0$, which is lost in the zero gap case.

In order to discuss the effects of XC interactions within the TF model, we note that density dependent expressions for both the
exchange and correlation energy per electron in graphene, are available in the LDA only for density variations with respect to
the equilibrium case of intrinsic, or undoped graphene having $\mu=0$, in the limits of zero temperature, zero gap, and
linearized band DOS.\cite{Rossi_2008,Polini_2008} Therefore, we specialize Eq.\ (\ref{integral}) to those parameters and convert
it to an integral equation for the potential energy $U(\vx)=e\phi(\vx)$,
\begin{eqnarray}
U(r)&=&\frac{e^2}{\ebg^0}\frac{Z}{\sqrt{r^2+z_0^2}}-V_\mathrm{xc}\left(n\left(U(r)\right)\right) \nonumber
\\
&-&4\frac{e^2}{\ebg^0} \int_0^\infty dr^\prime\,r^\prime\frac{n\left(U(r^\prime)\right)
}{r^\prime+r}\,K\!\left(\frac{2\sqrt{r^\prime r}}{r^\prime+r}\right), \label{TFD}
\end{eqnarray}
where $K$ is the complete elliptic integral of the first kind,\cite{Abramowitz} and the function $n(U)$ is obtained from Eq.\
(\ref{dilog}) in the limit of zero temperature as follows
\begin{eqnarray}
n(U)=\frac{U^2\,\mathrm{sgn}(U)}{\pi(\hbar v_F)^2}. \label{nzeroT}
\end{eqnarray}
For the XC potential energy, $V_\mathrm{xc}(n)$, we use the expressions provided by Polini \textit{et al.}\cite{Polini_2008} in
their subsections II.\ A and B. As a reference, we quote here the dominant contribution to this energy at low densities, which
is on the order of
\begin{eqnarray}
V_\mathrm{xc}(n)\sim\frac{e^2}{\ebg^0}\sqrt{\vert n\vert}\,\ln\!\left(\frac{n_0}{\vert n\vert}\right)\,\mathrm{sgn}(n)
\label{Vx}
\end{eqnarray}
for $\vert n\vert \lll n_0=\eta\times0.7635$ \AA$^{-2}$, where $\eta$ is a cut-off parameter of the theory taking a value from
the interval $(0,1]$.\cite{Polini_2008} Finally, we note that Eq.\ (\ref{TFD}) with $V_\mathrm{xc}(n)$ set to zero appeared in
previous studies using the TF model.\cite{DiVicenzo_1984,Katsnelson_2006,Fogler_2007,Brey_2009}

\subsection{Linear models}

Going back to the TF integral equation, Eq.\ (\ref{integral}), if the total potential $\phi(\vx)$ in the plane of graphene may
be treated as a weak perturbation of the equilibrium carrier charge density, then one can linearize Eq.\ (\ref{sigma_gr}),
$\sigma_\mathrm{gr}(\vx)\approx -e^2\phi(\vx)\partial n(\mu)/\partial \mu $, with $n(\mu)$ defined in Eq.\ (\ref{density}). This
facilitates the use of the Fourier transform in solving Eq.\ (\ref{integral}) thereby giving an approximate expression for the
total potential
\begin{eqnarray}
 \widetilde{\phi}({\bf k})=\frac{\ebg(k)}{\ebg(k)+v_C(k)\Pi(k)} \,\widetilde{\phi}_0({\bf k}),
\label{linTF}
\end{eqnarray}
where
\begin{eqnarray}
\ebg(k)=\left(1-\frac{\epsilon_\mathrm{s}-1}{\epsilon_\mathrm{s}+1}\,\mathrm{e}^{-2kh}\right)^{-1} \label{ebg}
\end{eqnarray}
is the background dielectric constant due to substrate, $v_C(k)=2\pi e^2/k$, and $\Pi(k)$ is the polarization function of free
graphene, which is constant in the LTF model, given by $\Pi_\mathrm{TF}\equiv\partial n(\mu)/\partial \mu $. In Eq.\
(\ref{linTF}), one needs to use the Fourier transform of the potential in Eq.\ (\ref{phi_zero}), which is given by
\begin{equation}
\label{pot_zero} \widetilde{\phi}_0({\bf k})=\frac{2\pi Ze}{k}\left\{
\begin{array}{ll}
\frac{\mathrm{e}^{-kz_0}}{\ebg(k)}, &  z_0>0
\\
\mathrm{e}^{kz_0}+\left[\frac{1}{\ebg(k)}-1\right]\mathrm{e}^{-kz_0}, &  -h<z_0<0
\\
\frac{\mathrm{e}^{kz_0}}{\ebg^0}, &  z_0<-h
\end{array}
\right.
\end{equation}
where $\ebg^0\equiv\ebg(0)=\left(\epsilon_\mathrm{s}+1\right)/2$.

In the zero gap limit, one obtains from Eq.\ (\ref{linTF}) a more compact expression for the total potential in the LTF model,
\begin{eqnarray}
 \widetilde{\phi}({\bf k})=\frac{2\pi Ze}{k\ebg^0+q_s}\,
 \mathrm{e}^{-k\vert z_0\vert}, \label{TF_pot_zero}
\end{eqnarray}
where the inverse screening length of free graphene, $q_s=2\pi e^2\Pi_\mathrm{TF}$, is obtained from Eq.\ (\ref{dilog}) within
the linearized DOS as\cite{Ando_2006}
\begin{eqnarray}
q_s\approx\frac{2g_d e^2}{\beta\left(\hbar v_F\right)^2}\,\ln\left[2\cosh\left(\beta\mu/2\right)\right]. \label{TF_screen}
\end{eqnarray}
It is clear then that, at zero temperature, intrinsic graphene cannot screen external charges in the LTF model because
$q_s\rightarrow 0$.\cite{Ando_2006} On the other hand, when either $n\neq 0$ or $T>0$, the inverse Fourier transform of Eq.\
(\ref{TF_pot_zero}) gives a total potential with the asymptotic form\cite{Katsnelson_2006}
$\phi(r)\sim\left(Ze\ebg^0\right)/\left(q_s^2r^3\right)$ for $r\gg q_s^{-1}\gg \vert z_0\vert$, and with the limiting value at
the origin
$\phi(0)=\left[Ze/\left(\ebg^0z_0\right)\right]\left[1-\zeta\,\mathrm{e}^{\zeta}\,\mathrm{E}_1\left(\zeta\right)\right]$, where
$\zeta\equiv q_sz_0/\ebg^0$ and $\mathrm{E}_1$ is the exponential integral function.\cite{Abramowitz}

The expression (\ref{linTF}) can also be used to obtain the total potential based on the RPA model if one employs the
polarization function, which is obtained for the linearized DOS at non-zero temperature
as\cite{Ando_2006,Ramezanali_2009,Hwang_2009}
\begin{widetext}
\begin{eqnarray}
\Pi_\mathrm{RPA}(k)=\frac{g_d}{\pi\beta\left(\hbar v_F\right)^2} \left\{\ln\left[2\cosh\left(\beta\mu/2\right)\right]+\frac{\pi
k}{8q_t} - \frac{k}{2q_t}\int_0^1
du\,\sqrt{1-u^2}\left(\frac{1}{1+\mathrm{e}^{uk/q_t-\beta\mu}}+\frac{1}{1+\mathrm{e}^{uk/q_t+\beta\mu}}\right)\right\},
 \label{TRPA}
\end{eqnarray}
\end{widetext}
where we have defined a thermal inverse screening length by $q_t=2/(\beta\hbar v_F)$. Note that $\mu$, which is to be used in
Eq.\ (\ref{TRPA}), can be obtained from Eq.\ (\ref{dilog}) for any given temperature and equilibrium charge carrier density $n$.
In the zero temperature limit, $\mu\rightarrow\varepsilon_F$ where $\varepsilon_F=\hbar v_Fk_F\mathrm{sgn}(n)$ is the Fermi
energy with $k_F=\sqrt{\pi\vert n\vert}$ being the Fermi momentum in graphene with the equilibrium charge carrier density $n$,
so that one obtains from Eq.\ (\ref{TRPA})\cite{Wunsch_2006,Hwang_2007,Allison_2009}
\begin{eqnarray}
\Pi_\mathrm{RPA}(k)&=&\frac{g_dk_F}{2\pi\hbar v_F}\left\{1+\left[\frac{k}{4k_F}\arccos\left(2\frac{k_F}{k}\right)\right.\right.
\nonumber
\\
&-&\left.\left.\frac{1}{2}\sqrt{1-\left(2\frac{k_F}{k}\right)^2}\right]H(k-2k_F)\right\} , \label{RPA}
\end{eqnarray}
where $H$ is the Heaviside unit step function. Unlike the LTF case, we see that $\Pi_\mathrm{RPA}(k)=k/\left(4\hbar v_F\right)$
in intrinsic graphene at zero temperature. Since this is also the short wavelength limit of $\Pi_\mathrm{RPA}(k)$ when $n\neq
0$, one may assert that the RPA result will yield a value for the total potential that is reduced by an approximate factor of
$\left[1+\pi r_s/\left(2\ebg^0\right)\right]^{-1}$, where $r_s\equiv e^2/\left(\hbar v_F\right)\approx 2.2$, when compared to
the corresponding value from the LTF approach for $k_F\sqrt{r^2+z_0^2}\ll 1$ at zero temperature and zero gap. On the other
hand, one can expect that the total potential will exhibit Friedel oscillations for $k_Fr\gg 1$ due to non-analyticity of the
RPA polarization function (\ref{RPA}) at $k=2k_F$, which will be gradually dampened as the temperature
increases.\cite{Wunsch_2006}

\subsection{Image interaction}

Once the integral equation, Eq.\ (\ref{integral}), is solved for the total potential in the plane of graphene, one can use Eq.\
(\ref{sigma_gr}) to evaluate the induced charge density in graphene, whose Fourier transform may be used in Eq.\ (\ref{ind}) to
yield the total induced potential for any value of $z$. This may be then used to calculate the nonlinear image force on the
external charge from the definition
\begin{eqnarray}
F_\mathrm{im}(z_0)=-Ze\left.\frac{\partial}{\partial z}\Phiind(\vx,z)\right\vert_{\vx={\bf 0},z=z_0}. \label{image}
\end{eqnarray}
Once the $z_0$ dependence of the image force is determined, the corresponding image potential may be obtained from the
definition $V_\mathrm{im}(z_0)=\int_{z_0}^\infty dz_0^\prime\,F_\mathrm{im}(z_0^\prime)$. While in the nonlinear TF case this
integration has to be performed numerically, in a linear theory one may use instead the usual definition of image potential as a
classical self-energy,\cite{Mowbray_2006} $V_\mathrm{im}(z_0)=\frac{1}{2}Ze\Phiind(\vx\!=\!{\bf 0},z\!=\!z_0)$, which gives for
$z_0>0$\cite{Allison_2009}
\begin{eqnarray}
V_\mathrm{im}(z_0)=\frac{\left(Ze\right)^2}{2}\int_0^\infty dk\,\mathrm{e}^{-2k
z_0}\left[\frac{1}{\ebg(k)+v_C(k)\Pi(k)}-1\right]. \label{image}
\end{eqnarray}

By using the LTF model, where $v_C(k)\Pi_\mathrm{TF}=q_s/k$, one obtains in the zero gap case
\begin{eqnarray}
 V_\mathrm{im}(z_0)=\frac{\left(Ze\right)^2}{4z_0\ebg^0}
\left[1-\ebg^0-2\zeta\,\mathrm{e}^{2\zeta}\,\mathrm{E}_1\left(2\zeta\right)\right],
 \label{TF_image}
\end{eqnarray}
where $\zeta\equiv q_sz_0/\ebg^0$. It is worthwhile mentioning that this expression gives asymptotically
$V_\mathrm{im}\sim-\left(Ze\right)^2\left[1/\left(4z_0\right)-1/\left(8q_s z_0^2\right)\right]$ for a heavily doped graphene
and/or sufficiently large distance, such that $q_sz_0\gg 1$. On the other hand, in the opposite limit, $q_sz_0\ll 1$, one finds
to the leading order $V_\mathrm{im}\sim\left(Ze\right)^2\left(1/\ebg^0-1\right)/\left(4z_0\right)$, as if graphene were totally
absent. When the RPA polarization function at zero temperature, Eq.\ (\ref{RPA}), is used in (\ref{image}) in the zero gap case,
one can show that similar limiting forms of the image potential exist, except that the effective background dielectric constant,
$\ebg^0$, is to be replaced by $\ebg^0+\pi r_s/2\approx \ebg^0+3.44$ when $k_Fz_0\ll 1$.\cite{Allison_2009}

\section{Results}

We first consider the case of a positive charge with $Z=1$ a distance 2 \AA\ above graphene lying on an SiO$_2$ substrate
($\epsilon_\mathrm{s}=3.9$) with several gap heights $h$, several equilibrium charge carrier densities $n$, and zero
temperature. This situation may be representative of a Li atom adsorbed atop supported graphene, where the effective charge
transfer is found to be around $Z= 0.9$, whereas the local DOS exhibits a resonant feature at about 0.9 eV above the neutrality
point of graphene's $\pi$ electron band due to hybridization with lithium's 2$s$ orbital.\cite{Chan_2008} Besides undoped
graphene with $n=0$, which was studied previously,\cite{DiVicenzo_1984,Katsnelson_2006,Fogler_2007} we also analyze the cases of
both electron ($n>0$) and hole ($n<0$) doping of graphene by a gate potential, making sure that the Fermi level stays well below
any chemisorption resonances in graphene's DOS ($n\lesssim 10^{13}$ cm$^{-2}$ for Li atom.\cite{Chan_2008})

\onecolumngrid
\begin{widetext}
\begin{figure}[t]
\centering
\includegraphics[width=1.1\textwidth]{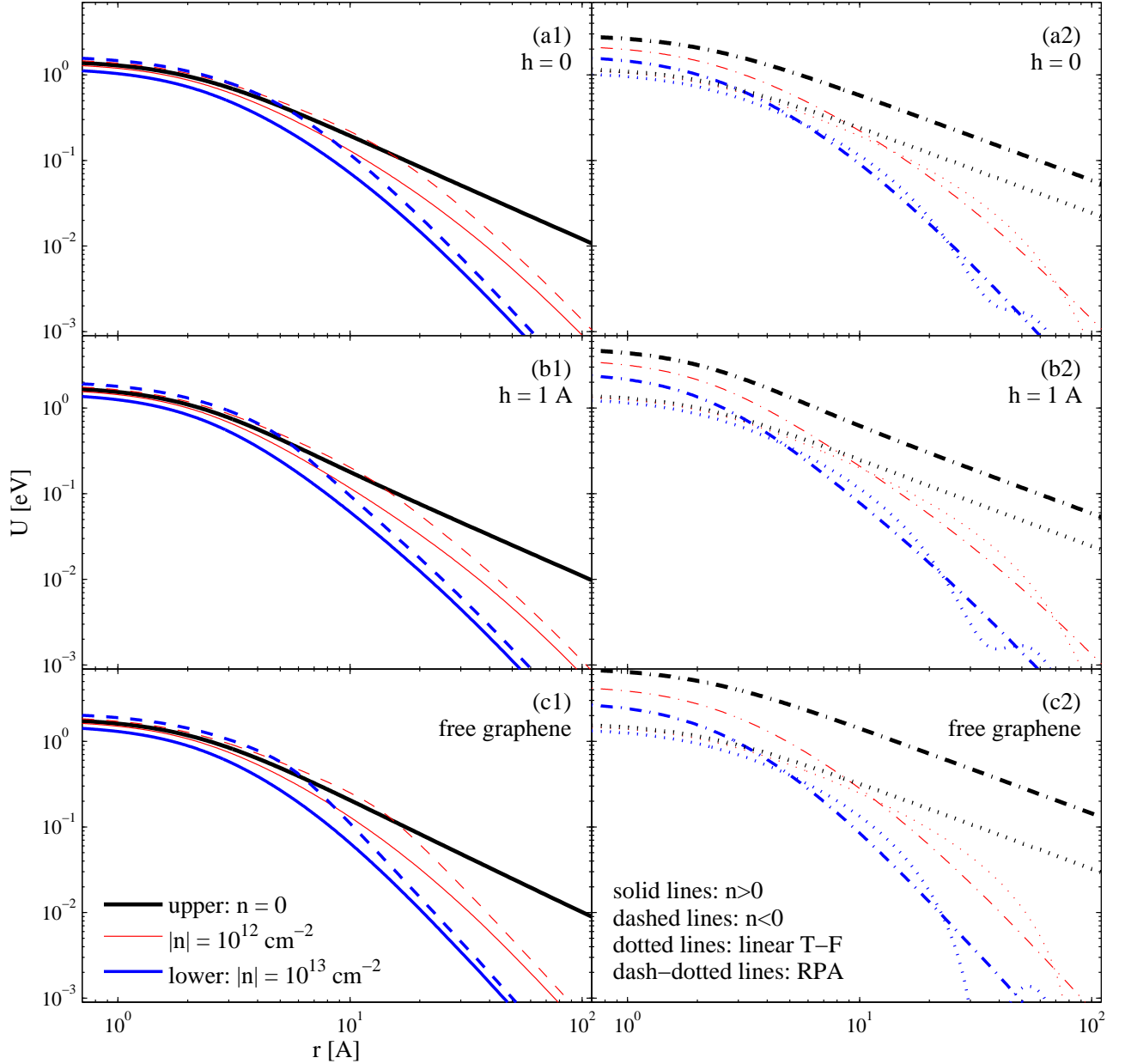}
\caption{
 The potential energy, $U(r)=e\phi(r)$ (in eV), due to an external proton at distance $z_0=2$ \AA\ above graphene at
zero temperature, as a function of the radial distance $r$ (in \AA) in the plane of graphene lying on an SiO$_2$ substrate with
the gap heights $h=0$ (panels a), 1 \AA\ (panels b), and $\infty$ (free graphene, panels c). Results from the nonlinear TF model
are shown in the column 1 for equilibrium densities $n=0$ (upper thick [black] solid line), $\pm 10^{12}$ (thin [red] solid and
dashed lines, respectively), and $\pm 10^{13}$ cm$^{-2}$ (lower thick [blue] solid and dashed lines, respectively). Results from
the linearized TF model and the RPA model are shown, respectively, by dash-dotted and dotted lines in the column 2 for densities
$\vert n\vert=0$ (upper thick [black] lines), $10^{12}$ (thin [red] lines), and $10^{13}$ cm$^{-2}$ (lower thick [blue] lines).
}
\end{figure}
\end{widetext}
\twocolumngrid

In Fig.\ 1 we show in the left column (1) the results for the potential energy $U(r)=e\phi(r)$, with $\phi(r)$ obtained from the
nonlinear TF equation Eq.\ (\ref{integral}) at zero temperature for $n=0$ (the upper thick solid line), $\pm 10^{12}$ (thin
solid and dashed lines, respectively), and $\pm 10^{13}$ cm$^{-2}$ (the lower thick solid and thick dashed lines, respectively),
and with $h=0$ (panels a), 1 \AA\ (panels b), and $\infty$ (i.e., free graphene, panels c). For the purpose of comparison, we
also show in the right column (2) of Fig.\ 1 the corresponding results obtained from both the LTF (dash-dotted lines) and the
RPA (dotted lines) models (with the line thicknesses matching those in the left column), with $\phi(r)$ calculated from Eq.\
(\ref{linTF}) using the appropriate polarization functions at zero temperature. [As a reference, note that, for free graphene,
the LTF result with $n=0$ actually shows the value of the unscreened potential in the plane of graphene, $U_0(r)=e\phi_0(r)$
with $\phi_0(r)$ given in Eq.\ (\ref{phi_zero}), whereas the corresponding RPA result shows that same potential reduced by the
dielectric constant of intrinsic graphene, $1+\pi r_s/2\approx 4.44$.] One can see in Fig.\ 1 that the main effects on the
potential come from increasing the doping density $\vert n\vert$. While all models exhibit strong variation with $n$ at large
distances $r$, one notices that both the nonlinear TF and the RPA results are surprisingly concentrated in a relatively narrow
range of values for the potential at short distances for all densities $n$. This seems to corroborate conclusions from a DFT
study that the induced density variations in graphene seem to saturate with increasing level of doping.\cite{Polini_2008}

While the LTF model appears to be a rather poor approximation to the nonlinear TF results at short distances $r$, their
agreement improves at large distances with increasing density $\vert n\vert$, as expected. Most strikingly, the RPA model gives
a surprisingly good approximation to the nonlinear TF results at short distances for all densities $n$, while exhibiting Friedel
oscillations around the LTF results at large distances for $n\neq 0$, with wavelengths that obviously scale with
$k_F^{-1}$.\cite{Wunsch_2006} However, for $n=0$, one sees an increasing disagreement between the nonlinear TF and the RPA
models with increasing distance, which may be attributed to a poor performance of the TF model in intrinsic graphene for induced
charge carrier densities below $10^{11}$ cm$^{-2}$, as suggested recently by Brey and Fertig.\cite{Brey_2009} On the other hand,
the TF model presumably gives a correct order of magnitude for nonlinear effects, if any, when the doping density $\vert n\vert$
increases, which are best seen by analyzing the effect of changing the sign of $n$ (equivalently, the sign of $Z$), because
linear models are insensitive to this sign. In that respect, one can clearly notice in the left column of Fig.\ 1 differences
between the potentials $U_+(r)$ for $n>0$ and $U_-(r)$ for $n<0$ in the nonlinear TF model, which will be further discussed in
Fig.\ 3 below.

Finally, one notices in Fig.\ 1 that, while the presence of a finite gap between graphene and substrate does not affect
qualitative behavior of the results, its quantitative effects may not be neglected in the values of the potential for all
densities shown. While this is particularly obvious at short distances for the nonlinear TF results, it is also interesting to
see how Friedel oscillations in the RPA model increase in amplitude with increasing gap. In fact, we have found that the RPA
potential can even change its sign at large distances $r$ for free graphene with large enough $\vert n\vert$. Given that the
size of gap is poorly defined parameter, with a plausible value of around $h=1$ \AA,\cite{Chan_2008,Khomyakov_2009} one should
be aware of its role in the total potential in graphene due to external charges.

\onecolumngrid
\begin{widetext}
\begin{figure}[h]
\centering
\includegraphics[width=1.1\textwidth]{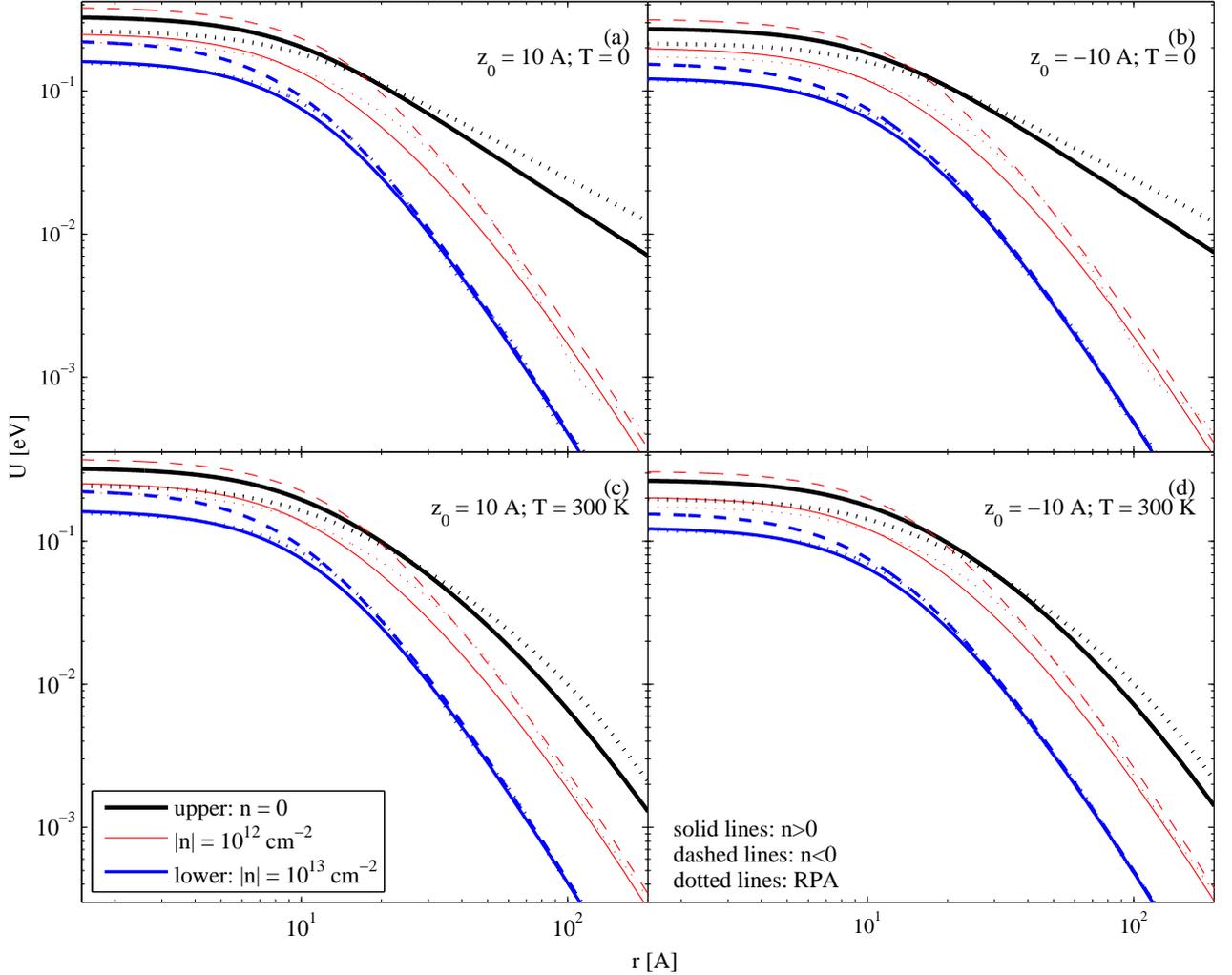}
\caption{ The potential energy, $U(r)=e\phi(r)$ (in eV), due to an external proton at distances $z_0=\pm 10$ \AA\ (left and
right columns, respectively) from graphene at $T=0$ (top row) and $T=300$ K (bottom row), as a function of the radial distance
$r$ (in \AA) in the plane of graphene lying on an SiO$_2$ substrate with the gap height $h=1$ \AA. Results from the nonlinear TF
model are shown for equilibrium densities $n=0$ (upper thick [black] solid line), $\pm 10^{12}$ (thin [red] solid and dashed
lines, respectively), and $\pm 10^{13}$ cm$^{-2}$ (lower thick [blue] solid and dashed lines, respectively). Results from the
RPA model are shown by dotted lines for densities $\vert n\vert=0$ (upper thick [black] line), $10^{12}$ (thin [red] line), and
$10^{13}$ cm$^{-2}$ (lower thick [blue] line). }
\end{figure}
\end{widetext}
\twocolumngrid

We next consider in Fig.\ 2 graphene on an SiO$_2$ substrate with the gap $h=1$ \AA, both at zero (panels a and b) and room
($T=300$ K, panels c and d) temperatures, with a charge $Z=1$ placed at larger distances of $z_0=\pm 10$ \AA\ away from
graphene. With $z_0=10$ \AA\ (panels a and c) we can represent a distant charge above graphene, such as a slowly moving
ion,\cite{Allison_2009} or an electron in an image-potential state,\cite{Silkin_2009} whereas the case $z_0=-10$ \AA\ (panels b
and d) represents a technologically relevant case of a charged impurity trapped deep in the SiO$_2$
substrate.\cite{Wu_2004,Rezende_2009} We compare the nonlinear TF results with those from the RPA model for $\vert n\vert=0$,
$10^{12}$, and $10^{13}$ cm$^{-2}$, shown with the same line styles and thicknesses as in Fig.\ 1. While the RPA results seem to
be quite close, apart from the Friedel oscillations, to those of the nonlinear TF model for $n>0$, the agreement between those
two models seems to have worsened at short distances for $n=0$ when compared to Fig.\ 1, which may have to do with the
problematic performance of the nonlinear TF model in intrinsic graphene exposed to weak perturbations, as mentioned
previously.\cite{Brey_2009}

On the other hand, one notices in Fig.\ 2 a much greater spread in the relative magnitudes of the potential at short distances
than in Fig.\ 1. This is partly due to the effect of doping in the presence of a much weaker external perturbation in Fig.\ 2
than in Fig.\ 1, so that the induced density variations involved in the results in Fig.\ 2 have not reached the effect of
saturation mentioned earlier.\cite{Polini_2008} An another cause for a larger spread of the potential at short distances in
Fig.\ 2 comes from the nonlinear effects, which will be further discussed in Fig.\ 3.

As regards the effect of finite temperature, one notices that its main role is to dampen the potential in intrinsic graphene at
distances $r\gtrsim 10$ \AA, both in the nonlinear TF and the RPA cases. This can be explained by assessing the TF inverse
screening length in Eq.\ (\ref{TF_screen}) in the zero density and the zero temperature limits, giving $q_s\rightarrow 4r_s
q_t\ln2$ and $q_s\rightarrow 4r_sk_F$, respectively. Therefore, one may conclude that screening of the potential at large
distances due to a non-zero temperature will prevail only for low enough charge-carrier densities, such that $\vert
n\vert<\left[2\ln2 k_BT/\left(\hbar v_F\right)\right]^2/\pi \approx 10^{11}$ cm$^{-2}$ at room temperatures. [In fact, we have
checked that nonlinear TF results for $\vert n\vert=10^{11}$ cm$^{-2}$ at zero temperature (not shown) are quite close to those
shown in Fig.\ 2 for intrinsic graphene at room temperature.] The effects of temperature on the nonlinearity of the potential
are further discussed in Fig.\ 3. On the other hand, while the Friedel oscillations are still visible in Fig.\ 2 in the RPA
results for zero temperature at large distances $r$ for $n\neq 0$, although they seem to be reduced in relative amplitude by the
increased distance $\vert z_0\vert$ when compared to the oscillations seen in Fig.\ 1, one notices that the increased
temperature dampens the Friedel oscillations in Fig.\ 2, as expected.

We note finally that, by analyzing the asymmetry in the results with respect to the change in sign of $z_0$ in Fig.\ 2, we yet
again emphasize the role of a finite gap, because all results would be independent of that sign in the zero gap case. It is
remarkable that the gap of only $h=1$ \AA\ affects not only the values of the potential ar short distances, but also the
magnitudes of the asymmetry in the nonlinear TF results with respect to the sign of $n\neq 0$ at short distances.

\begin{figure}[]
\centering
\includegraphics[width=0.52\textwidth]{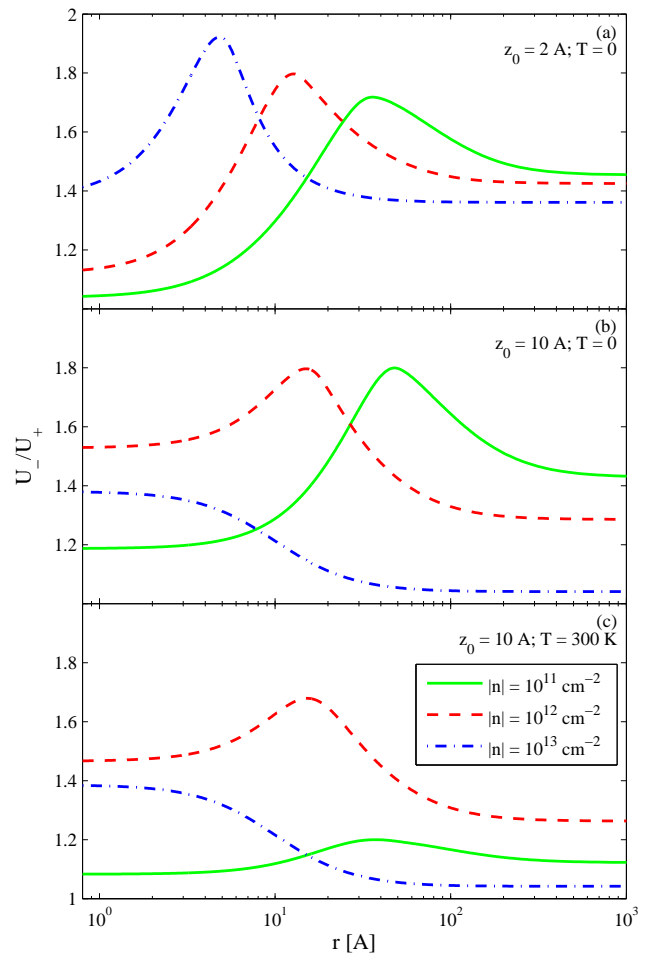}
\caption{The ratio $U_-(r)/U_+(r)$ of the nonlinear potential energies $U_-(r)$ and $U_+(r)$ corresponding to, respectively,
negative (hole doping) and positive (electron doping) signs of the equilibrium charge carrier densities $\vert n\vert=10^{11}$
(solid [green] lines), $10^{12}$ (dashed [red] lines), and $10^{13}$ cm$^{-2}$ (dash-dotted [blue] lines), is shown as a
function of the radial distance $r$ (in \AA) in the plane of graphene for a proton at distances $z_0=2$ \AA\ with $T=0$ (panel
a), $z_0=10$ \AA\ with $T=0$ (panel b), and  $z_0=10$ \AA\ with $T=300$ K (panel c), above graphene lying on an SiO$_2$
substrate with the gap $h=1$ \AA.}
\end{figure}
Nonlinear effects in screening of an external charge by doped graphene, seen in Figs.\ 1(b1) and 2(a), are summarized in Fig.\
3, with the inclusion of the results for doping density of $\vert n\vert=10^{11}$ cm$^{-2}$. We show the ratio $U_-(r)/U_+(r)$
of the potential energies $U_-(r)$ and $U_+(r)$, which are obtained from Eq.\ (\ref{integral}) with, respectively, negative
(hole doping) and positive (electron doping) signs of densities $\vert n\vert=10^{11}$ (solid lines), $10^{12}$ (dashed lines),
and $10^{13}$ cm$^{-2}$ (dash-dotted lines), for a charge $Z=1$ at two distances with two temperatures: $z_0=2$ \AA\ and $T=0$
(panel a), $z_0=10$ \AA\ and $T=0$ (panel b), and $z_0=10$ \AA\ and $T=300$ K (panel c), for graphene lying on an SiO$_2$
substrate with the gap $h=1$ \AA. One notices in Fig.\ 3 that the ratio $U_-(r)/U_+(r)$ may reach quite large values (up to a
factor of two), indicating that nonlinear effects in screening of external charges may be very strong. In particular, this ratio
reaches maximum values at certain distances $r_c$ that obviously depend on both the doping density $\vert n\vert$ and the
strength of external perturbation determined by $z_0$. [We note that the difference $U_-(r)-U_+(r)$ is always found to peak at
the origin, $r=0$.]

The maxima in the ratios, seen in Fig.\ 3, may be explained by the fact that, for the hole doping ($n<0$) of graphene in the
presence of a positive external charge, there will be a local re-doping with electrons, or discharging of graphene, giving rise
to a local shift of the $\pi$ electron band DOS, such that the condition $U_-(r_c)\approx \hbar v_Fk_F$ may be reached,
indicating that the Fermi level is pushed back to cross the neutrality point at some distance $r=r_c$. Since there are fewer
states available in the DOS around the neutrality point, the screening ability of graphene is reduced around $r=r_c$ when $n<0$,
resulting in a higher value of the total potential than in the case of electron doping ($n>0$), so that one may expect that an
inequality $U_-(r)>U_+(r)>0$ will hold for a range of distances $r$ around $r_c$. For example, in Fig.\ 3(a), the external
charge is so close to graphene at zero temperature that it provides a strong enough perturbation, giving rise to the local
discharging for all three doping densities, $\vert n\vert=10^{11}$, $10^{12}$, and $10^{13}$ cm$^{-2}$, so that three maxima in
the ratio $U_-(r)/U_+(r)$ occur around distances $r_c\approx 35.6$, 12.7, and 4.8 \AA, respectively. The corresponding values of
the potential $U_-(r_c)$ at these distances are found to be 0.037, 0.137, and 0.495 eV, respectively, which scale reasonably
close to the Fermi level shift at the three doping densities, $\vert\varepsilon_F\vert=\hbar v_Fk_F\approx$ 0.037, 0.117, and
0.368 eV.

On the other hand, when the charge is removed to distance $z_0=10$ \AA\ at zero temperature in Fig.\ 3(b), the perturbation is
still strong enough to discharge graphene for the two lower doping densities [with the peaks occurring at similar distances,
$r_c\approx 47.8$ and 15.0 \AA, and with similar potential values, $U_-(r_c)\approx 0.041$ and 0.153 eV, as in Fig.\ 3(a)], but
is not sufficient do force the Fermi level to cross the neutrality point for the highest density of $\vert n\vert=10^{13}$
cm$^{-2}$, for which a maximal local discharging of graphene occurs directly underneath the external charge. Furthermore, when
the temperature is raised to $T=300$ K for $z_0=10$ \AA, the ratio $U_-(r)/U_+(r)$ for the two higher doping densities is barely
affected, but the ratio for the lowest density of $\vert n\vert=10^{11}$ cm$^{-2}$ appears to be largely suppressed in Fig.\
3(c) as compared to Fig.\ 3(b). One can still see a maximum in this ratio around a distance similar to that in Fig.\ 3(b), i.e.,
$r_c\approx 37.2$ \AA\ with $U_-\approx 0.045$ eV, but the peak value of the ratio $U_-(r)/U_+(r)$ for  $\vert n\vert=10^{11}$
cm$^{-2}$ has dropped from about 1.8 for $T=0$ K to about 1.2 for $T=300$ K. While the results in Fig.\ 3(c) confirm the
conclusion drawn from Fig.\ 2 that, at room temperature, the screening ability of graphene is affected for sufficiently low
doping densities, such that $\vert n\vert\lesssim 10^{11}$ cm$^{-2}$, it is now clear that the role of elevated temperature,
when it prevails the effects of doping density, is to the suppress the nonlinear effects.

\begin{figure}
\centering
\includegraphics[width=0.52\textwidth]{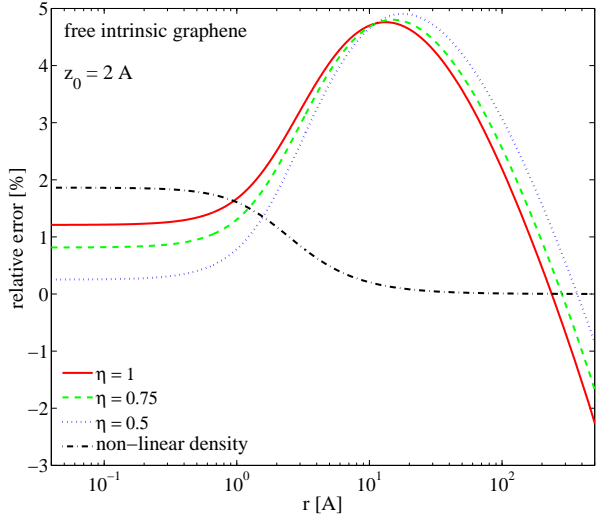}
\caption{The relative error in the potential energy, $U(r)=e\phi(r)$ (in \%), from the nonlinear TF model for a proton at
distance $z_0=2$ \AA\ from free, intrinsic ($n=0$) graphene at zero temperature, due to the inclusion of the exchange and
correlation energy \cite{Polini_2008} with values of the cutoff parameter being $\eta=1$ (solid [red] line), 0.75 (dashed
[green] line), and 0.5 (dotted [blue] line), as well as due to the nonlinear correction to graphene's $\pi$ electron band
density of state (dash-doted [black] line).}
\end{figure}
All results shown in Figs.\ 1-3 were obtained by taking into account in Eq.\ (\ref{density}) the effects of nonlinearity in the
band DOS of graphene, $\rho(\varepsilon)$, because we suspected that the value of the potential $U(r)$ may exceed locally (that
is, directly underneath the external charge) the cutoff value of about 1 eV that validates the linear approximation for
$\rho(\varepsilon)$. Our calculations show that the effect of this nonlinearity is relatively weak, giving corrections up to
several percent for distances $\vert z_0\vert>1.5$ \AA. This is illustrated in Fig.\ 4 for free, intrinsic ($\mu=0$) graphene at
zero temperature with a charge $Z=1$ placed at $z_0=2$ \AA, where we show by the dash-dotted line the relative error in the
total potential when Eq.\ (\ref{integral}) is solved with density $n$ from Eq.\ (\ref{dilog}) and from Eq.\ (\ref{density}) with
a nonlinear DOS $\rho(\varepsilon)$.\cite{Neto_2009} One can see that the peak error of about 2 \% occurs at the origin and
diminishes at distances greater than a few \AA.

We further estimate the effects of the exchange and correlation interactions, which have been neglected so far in solving the
nonlinear TF equation (\ref{integral}). We use the expression $V_\mathrm{xc}(n)$ for the XC potential energy given by Polini
\textit{at al.}\cite{Polini_2008} in the LDA and, since the formalism providing $V_\mathrm{xc}(n)$ is restricted to intrinsic
graphene at zero temperature within the linear approximation for $\rho(\varepsilon)$,\cite{Polini_2008} we solve the nonlinear
equation, Eq.\ (\ref{TFD}) with Eq.\ (\ref{nzeroT}), for free graphene ($\ebg^0=1$) with the charge $Z=1$ a distance $z_0=2$
\AA\ away. The result is compared to the solution when $V_\mathrm{xc}$ is set to zero by showing in Fig.\ 4 the relative error
of such a comparison for several values of the cutoff parameter $\eta$.\cite{Polini_2008} One can see in Fig.\ 4 that the
relative error due to the XC interactions is relatively small at short distances $r$, and is comparable to the error due to the
nonlinear band DOS. However, the error due to the XC interactions increases and reaches a maximum of about 5\% at distances on
the order of $r=10$ \AA\ or more, reverses its sign at still greater distances of about $r=100$ \AA\ or more, and presumably
continues growing further in magnitude. While this is a relatively small error at radial distances where the total potential has
a significant value, we note that the error due to the XC interaction may be larger when external charge is placed further away
from graphene, as noted earlier.\cite{Polini_2008} However, because of the limitation of the theory for XC interactions to local
perturbations of charge carrier density relative to intrinsic graphene at zero temperature,\cite{Rossi_2008,Polini_2008} we no
longer pursue the analysis of the XC effects in our nonlinear TF approach.

\begin{figure}
\centering
\includegraphics[width=0.52\textwidth]{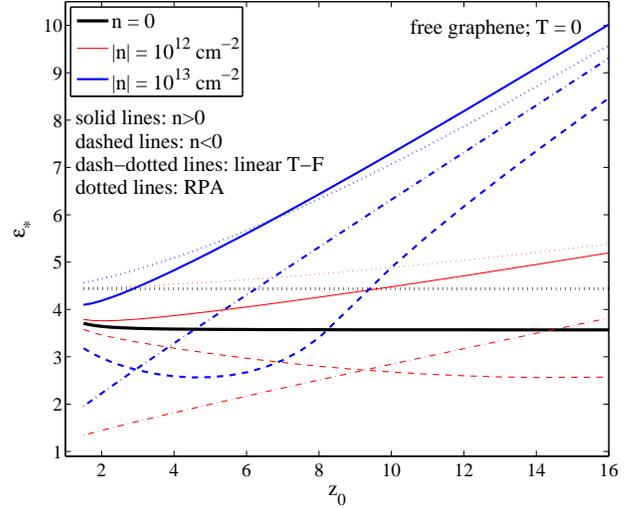}
\caption{The effective dielectric constant $\estar$ in the image force, written as $F_\mathrm{im}=(Ze/2z_0)^2(1/\estar-1)$, as a
function of distance $z_0$ (in \AA) for a proton above free graphene at zero temperature. Results from the nonlinear TF model
are shown for equilibrium densities $n=0$ (lower thick [black] solid line), $\pm 10^{12}$ (thin [red] solid and dashed lines,
respectively), and $\pm 10^{13}$ cm$^{-2}$ (upper thick [blue] solid and dashed lines, respectively). Results from the
linearized TF model and the RPA model are shown, respectively, by dash-dotted and dotted lines for densities $\vert n\vert=0$
(lower thick [black] lines), $10^{12}$ (thin [red] lines), and $10^{13}$ cm$^{-2}$ (upper thick [blue] lines).}
\end{figure}
While the results in Figs.\ 1-4 elucidate local properties of the solution of the nonlinear TF equation, Eq.\ (\ref{integral}),
we now turn to analyzing the image force $F_\mathrm{im}$ on a point charge as a quantity that provides an integrated information
on the effects of doping and nonlinear screening in graphene. We first consider free graphene at zero temperature, and represent
the nonlinear image force in the form reminiscent of the classical image force of a point charge $Ze$ in vacuum, a distance
$z_0$ away from a layer of dielectric material with an effective dielectric constant $\estar$, given by
\begin{eqnarray}
 F_\mathrm{im}=\frac{\left(Ze\right)^2}{4z_0^2}\left[\frac{1}{\estar(z_0)}-1\right]. \label{Fimstar}
\end{eqnarray}
In this way, the $z_0$ dependent parameter $\estar$ provides a measure of the polarizability of free graphene. We use the same
line styles and thicknesses as in Fig.\ 1 to show in Fig.\ 5 the results of the nonlinear TF calculations of $\estar$ as a
function of $z_0$ for $\vert n\vert=0$, $10^{12}$, and $10^{13}$ cm$^{-2}$, along with the corresponding LTF and RPA results
obtained from Eq.\ (\ref{image}) with an appropriate polarization function by taking the derivative,
$F_\mathrm{im}=-dV_\mathrm{im}/dz_0$. One can see in Fig.\ 5 a strong dependence of the nonlinear TF image force on both the
magnitude and the sign of charge carrier density $n$, whereas the linear results seem to work only at large enough distances
$z_0$, with the RPA model showing a better agreement with the nonlinear TF results than the TF model. Notice that the slopes of
the LTF lines follow from taking the derivative of the asymptotic limit of the image potential in Eq.\ (\ref{TF_image}), and are
given for $n\neq 0$ by the zero temperature limit of the inverse screening length in Eq.\ (\ref{TF_screen}), $q_s=4r_sk_F$. On
the other hand, the nearly horizontal lines for the nonlinear TF and the RPA models with $n=0$ show that intrinsic graphene
behaves as a layer of material with effective dielectric constants of $\approx 3.57$ and $\approx 1+\pi r_s/2\approx 4.44$,
respectively.

\begin{figure}
\centering
\includegraphics[width=0.52\textwidth]{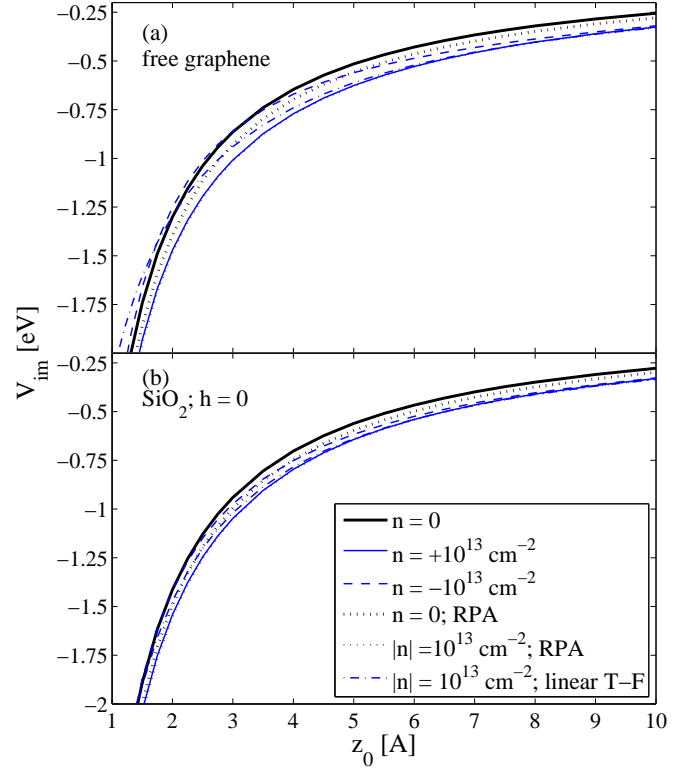}
\caption{ Image potential $V_\mathrm{im}$ (in eV) of a proton as a function of its distance $z_0$ (in \AA) above graphene at
zero temperature, for free graphene (upper panel) and for graphene lying on an SiO$_2$ substrate with zero gap (lower panel).
Results from the nonlinear TF model are shown for equilibrium densities $n=0$ (upper thick [black] solid line) and $\pm 10^{13}$
cm$^{-2}$ (lower thin [blue] solid and dashed lines, respectively). Results from the RPA model are shown by dotted lines for
densities $\vert n\vert=0$ (upper thick [black] line) and $10^{13}$ cm$^{-2}$ (lower thin [blue] line), as well as from the
linearized TF model for density $\vert n\vert=10^{13}$ cm$^{-2}$ (thin [blue] dash-dotted line). }
\end{figure}

Finally, we analyze in Fig.\ 6 the image potential on a point charge $Z=1$ above free graphene (panel a) and in the presence of
a SiO$_2$ substrate with zero gap (panel b), at zero temperature. We show the results due to the nonlinear TF and the RPA models
for $n=0$ (thick solid and dotted lines, respectively) and $\pm 10^{13}$ cm$^{-2}$ (thin solid and dashed lines for the
nonlinear TF, and thin dotted line for the RPA model), as well as the results due to the LTF model for $\vert n\vert=10^{13}$
cm$^{-2}$ (thin dash-dotted line). We note that the nonlinear results were obtained by integrating the corresponding image force
from $z_0$ up to typically 400 \AA. One notices a relatively close grouping of all results, indicating that the linear models
provide good approximations, especially at high density and large distances $z_0$.

However, the effects of doping of graphene are seen to be still quite strong giving, e.g., in the nonlinear TF model for free
graphene the image potential of $V_\mathrm{im}\approx-0.32$ eV at $z_0=10$ \AA\ when $\vert n\vert=10^{13}$ cm$^{-2}$, as
opposed to $V_\mathrm{im}\approx-0.26$ eV found at the same distance above intrinsic graphene. This points to possibly strong
effects of doping in the asymptotic region of distances of relevance to the image potential states.\cite{Silkin_2009} While the
discrepancy between the RPA and the nonlinear TF results, seen in Fig.\ 6 for free graphene at zero doping, stems from the
difference seen in Fig.\ 5 between the effective dielectric constants of intrinsic graphene in those two models, one notices a
near-perfect agreement of the RPA model with the nonlinear TF model in graphene doped by electrons to $n=10^{13}$ cm$^{-2}$.
However, nonlinear effects are still quite strong, especially at short distances, as illustrated by the observed asymmetry in
the nonlinear TF model with respect to the sign of $n\neq 0$. For example, one finds in Fig.\ 6(a) that the image potential
takes the value of $V_\mathrm{im}\approx -1.93$ eV at $z_0\approx 1.5$ \AA\ above free graphene with $n=10^{13}$ cm$^{-2}$, as
opposed to $V_\mathrm{im}\approx -1.66$ eV at the same distance with $n=-10^{13}$ cm$^{-2}$. This asymmetry due to doping of
graphene by electrons or holes may have interesting and important consequences for, e.g., chemisorption of a Li atom, where the
image potential shift of its 2$s$ orbital level may be controlled by the applied gate potential and used to move around the
resonance in the local DOS, and even possibly break the ionic bond between the Li atom and graphene. Finally, we note that we
have estimated numerically the effects of non-zero temperature and the XC interactions in the nonlinear image potential for
intrinsic graphene, and found that both these effects are negligible compared to the above effects of the doping density and
nonlinear screening.

\section{Concluding Remarks}

We have solved a nonlinear TF equation for the radial dependence of electric potential in the plane of single-layer graphene due
to an external point charge in the presence of a dielectric substrate with a finite graphene-substrate gap, $h$, paying special
attention to the effects of equilibrium charge carrier density $n$, temperature $T$, and separation between the charge and
graphene $\vert z_0\vert$. Large effects were found due to variations in both the magnitude and the sign of $n$, illustrating
the importance of both doping of graphene and the nonlinear screening, respectively. Temperature was found to mostly affect
screening at low doping densities, satisfying the inequality $k_F=\sqrt{\pi\vert n\vert}\lesssim k_BT/\left(\hbar v_F\right)$,
in such a way as to suppress the nonlinear effects. In addition, the existence of a non-zero gap, $h$, between the substrate and
graphene was found to exert non-negligible effects on the potential, mostly at short radial distances. We have moreover analyzed
the effects in the potential due to nonlinear corrections in the density of states of graphene's $\pi$ electron bands, as well
as due to the exchange and correlation interactions for the case of free, intrinsic graphene at zero temperature. While the
former effect gives corrections of up to a few percent at positions directly underneath the external charge and diminishes at
distances further out, the latter effect gives rise to the corrections of up to 5 \% at intermediate and large radial distances.

Comparisons were made with the results from a linearized TF (LTF) equation and from the RPA model of dielectric screening in
graphene. While the LTF results are generally close to the nonlinear TF results at large radial distances and high densities
$\vert n\vert$ only, the RPA model also exhibits an improved agreement with the nonlinear TF model at short radial distances,
owing to the short wavelength dielectric constant of graphene, which results from the inter-band electron transitions captured
by the RPA model.\cite{Hwang_2007,Polini_2008} Unlike the TF models, the RPA results exhibit Friedel oscillations around the
potential from the linearized TF model at large radial distances in doped graphene, with amplitudes that increase with
increasing gap $h$, but are dampened by increasing separation $\vert z_0\vert$ and increasing temperature.

Our most important conclusion is that nonlinear effects are strong over a broad range of radial distances, even at high doping
densities $\vert n\vert$ and large separations $\vert z_0\vert$, as illustrated by the large ratios of the potential evaluated
from the nonlinear TF model with the same amounts of doping by holes ($n<0$) and by electrons ($n>0$). This may be explained by
a local shift of graphene's density of states so that the Fermi level is forced to cross the neutrality point in that density at
a certain radial distance, thereby reducing graphene's polarizability when doping occurs with carriers of the same charge sign
as the external particle. This asymmetry in the scattering potential for charge carriers in graphene with respect to the sign of
$n$ may be responsible for the observed asymmetry in graphene's conductivity as the sign of the gate potential
changes.\cite{Farmer_2009} However, such an effect of nonlinear screening of external charges will be suppressed at low doping
densities when the temperature is sufficiently elevated, as described above.

Finally, we have analyzed the image interaction of an external charge due to polarization of graphene, where we compared the
results evaluated from the solution of the nonlinear TF equation with those from the LTF and the RPA models. After elucidating
the strong doping and nonlinear effects in the image force above free graphene at zero temperature, we have presented results
for an image potential obtained by numerical integration of the nonlinear image force up to large distances from graphene, and
compared them with the results of the linear models. The nonlinear image potential was found to exhibit relative variations due
to doping of graphene up to $\vert n\vert =10^{13}$ cm$^{-2}$, which may reach over 20 \% at distances $\vert z_0\vert\sim 10$
\AA, as well as due to the nonlinear screening, where relative variation with the sign of $n$ may reach about 20 \% at short
distances, on the order of $\vert z_0\vert\sim 1$ \AA. These variations in the image potential were found to be somewhat reduced
in the presence of an SiO$_2$ substrate.

Our results for the electric potential in the plane of graphene due to an external charge may be relevant for calculations of
its conductivity based on the Boltzmann transport model,\cite{Neto_2009,Adam_2009} where this potential may be used directly in
an expression for the transport relaxation time in the Born approximation, to reveal the effects of doping, nonlinear screening
and temperature on conductivity. While this task is left for a future contribution, we comment here that our nonlinear TF
results are likely to yield calculable effects due to the asymmetry in charge of the external particles,\cite{Farmer_2009} based
on the presently observed asymmetry with respect to the sign of $n$ for a positive external charge. Moreover, our results for
the nonlinear image potential may be found helpful in studying chemical processes near graphene, e.g., alkali atom chemisorption
and intercalation,\cite{Chan_2008} as well as in the recent work on the electron image-potential states near
graphene.\cite{Silkin_2009}

\begin{acknowledgments}
This work was supported by the Natural Sciences and Engineering Research Council of Canada.
\end{acknowledgments}


\begin{thebibliography}{10}

\bibitem{Novoselov_2004}
K. S. Novoselov, A. K. Geim, S. V. Morozov, D. Jiang, Y. Zhang, S. V. Dubonos, I. V. Grigorieva, and A. A. Firsov, Science
\textbf{306}, 666 (2004).

\bibitem{Neto_2009}
A. H. C. Neto, F. Guinea, N. M. R. Peres, K. S. Novoselov, and A. K. Geim, Rev. Mod. Phys. \textbf{81}, 109 (2009).

\bibitem{Tan_2007}
Y. W. Tan, Y. Zhang, K. Bolotin, Y. Zhao, S. Adam S, E. H. Hwang, S. Das Sarma, H. L. Stormer, and P. Kim, Phys. Rev. Lett.
\textbf{99}, 246803 (2007).

\bibitem{Chen_2008}
J. H. Chen, C. Jang, S. Adam, M. S. Fuhrer, E. D. Williams, and M. Ishigami, Nature Physics \textbf{4}, 377 (2008).

\bibitem{Das_2008}
A. Das, S. Pisana, B. Chakraborty, S. Piscanec, S. K. Saha, U. V. Waghmare, K. S. Novoselov, H. R. Krishnamurthy, A. K. Geim, A.
C. Ferrari, and A. K. Sood, Nature Nanotechnology \textbf{3}, 210 (2008).

\bibitem{Du_2008}
X. Du, I. Skachko, A. Barker, and E. Y. Andrei, Nature Nanotechnology \textbf{3}, 491 (2008).

\bibitem{Bolotin_2008}
K. I. Bolotin, K. J. Sikes, J. Hone, H. L. Stormer, and P. Kim, Phys. Rev. Lett. \textbf{101}, 096802 (2008).

\bibitem{Ando_2006}
T. Ando, J. Phys. Soc. Japan \textbf{75}, 074716 (2006).

\bibitem{Adam_2007}
S. Adam, E. H. Hwang, V. M. Galitski, and S. Das Sarma, Proc. Natl. Acad. Sci. U.S.A. \textbf{104}, 18392 (2007).

\bibitem{Adam_2009}
S. Adam, E. H. Hwang, E. Rossi, and S. Das Sarma, Solid State Communications \textbf{149}, 1072 (2009).

\bibitem{Shytov_2007}
A. V. Shytov, M. I. Katsnelson, and L. S. Levitov, Phys. Rev. Lett. \textbf{99}, 236801 (2007).

\bibitem{Novikov_2007}
D.S. Novikov, Phys. Rev. B \textbf{76}, 245435 (2007).

\bibitem{Pereira_2008}
V. M. Pereira, V. N. Kotov, and A. H. C. Neto, Phys. rev. B \textbf{78}, 085101 (2008).

\bibitem{Terekhov_2008}
I. S. Terekhov, A. I. Milstein, V. N. Kotov, and O. P. Sushkov, Phys. Rev. Lett. \textbf{100}, 076803 (2008).

\bibitem{DiVicenzo_1984}
D. P. DiVicenzo and E. J. Mele, Phys. Rev. \textbf{29}, 1685 (1984).

\bibitem{Katsnelson_2006}
M. I. Katsnelson, Phys. Rev. B \textbf{74}, 201401(R) (2006).

\bibitem{Fogler_2007}
M. M. Fogler, D. S. Novikov, and B. I. Shklovskii, Phys. Rev. B \textbf{76}, 233402 (2007).

\bibitem{Radovic_2008}
I. Radovic, L. J. Hadzievski, and Z. L. Miskovic, Phys. Rev. B \textbf{77}, 075428 (2008).

\bibitem{Rossi_2008}
E. Rossi and S. Das Sarma, Phys. Rev. Lett. \textbf{101}, 166803 (2008).

\bibitem{Polini_2008}
M. Polini, A. Tomadin, R. Asgari, and A. H.  MacDonald, Phys. Rev. B \textbf{78}, 115426 (2008).

\bibitem{Wu_2004}
Y. Wu and M. A. Shannon, J. Micromech. Microeng. \textbf{14}, 989 (2004).

\bibitem{Rezende_2009}
C.A. Rezende, R.F. Gouveia, M.A. da Silva, and F. Galembeck, J. Phys. - Cond. Matt.\textbf{ 21}, 263002 (2009)

\bibitem{Shedin_2007}
F. Schedin, A. K. Geim, S. V. Morozov, E. W. Hill, P. Blake, M. I. Katsnelson, and K.S. Novoselov, Nature Materials \textbf{6},
652 (2007).

\bibitem{Chan_2008}
k. T. Chan, J. B. Neaton, and M. L. Cohen, Phys. Rev. B \textbf{77}, 235430 (2008).

\bibitem{Neto_2009_a}
A. H. C. Neto, V. N. Kotov, J. Nilsson, V. M. Pereira, N. M. R. Peres, and B. Uchoa, Solid State Communications \textbf{149},
1094 (2009).

\bibitem{Algdal_2007}
J. Algdal, T. Balasubramanian, M. Breitholtz, T. Kihlgren, and L. Wallden, Surf. Sci. \textbf{601}, 1167 (2007).

\bibitem{Gumbs_2009}
G. Gumbs, D. Huang, and P.M. Echenique, Phys. Rev. B \textbf{79}, 035410 (2009).

\bibitem{Silkin_2009}
V.M. Silkin, J. Zhao, F. Guinea, E.V. Chulkov, P.M. Echenique, and H. Petek, Phys. Rev. B \textbf{80}, 121408 (2009).

\bibitem{Allison_2009}
K.F. Allison, D. Borka, I. Radovic, L. J. Hadzievski, and Z. L. Miskovic, Phys. Rev. B \textbf{80}, 195405 (2009).

\bibitem{Ishigami_2007}
M. Ishigami, J. H. Chen, W. G. Cullen, M. S. Fuhrer, and E. D. Williams, Nano Lett. \textbf{7}, 1643 (2007).

\bibitem{Fratini_2008}
S. Fratini and F. Guinea, Phys. Rev. B \textbf{77}, 195415 (2008).

\bibitem{Jang_2008}
C. Jang, S. Adam, J. H. Chen, D. Williams, S. Das Sarma, and M. S. Fuhrer, Phys. Rev. Lett. \textbf{101}, 146805 (2008).

\bibitem{Chen_2009}
F. Chen, J. L. Xia, and N. J. Tao, Nano Lett. \textbf{9}, 1621 (2009).

\bibitem{Chen_2009_a}
F. Chen, J. L. Xia, D. K. Ferry, and N. J. Tao, Nano Lett. \textbf{9}, 2571 (2009).

\bibitem{Wunsch_2006}
B. Wunsch, T. Stauber, F. Sols, and F. Guinea, New Journal of Physics \textbf{8}, 318 (2006).

\bibitem{Hwang_2007}
E.H. Hwang and S. Das Sarma, Phys. Rev. B \textbf{79}, 165404 (2007).

\bibitem{Brey_2009}
L. Brey mand H.A. Fertig, Phys. Rev. B \textbf{80},  035406  (2009).

\bibitem{Ramezanali_2009}
M.R. Ramezanali, M.M. Vazifeh, R. Asgari, M. Polini, and A. H.  MacDonald, J. Phys. A: Math. Theor. 42, 214015 (2009)

\bibitem{Hwang_2009}
E.H. Hwang and S. Das Sarma, Phys. Rev. B \textbf{75}, 205418 (2009).

\bibitem{Abramowitz}
M. Abramowitz and I. A. Stegun, \textit{Handbook of Mathematical Functions}, (National Bureau of Standards, Washington, 1965).

\bibitem{Doerr_2004}
T. P. Doerr and Y. K. Yu, Am. J. Phys. \textbf{72}, 190 (2004).

\bibitem{Mowbray_2006}
D. J. Mowbray, Z. L. Miskovic and F. O. Goodman, Phys. Rev. B \textbf{74}, 195435 (2006).

\bibitem{Mermin_1965}
N. D. Mermin, Phys. Rev. \textbf{137}, A1441 (1965).

\bibitem{Yonei_1987}
K. Yonei, J. Ozaki, and Y. Tomishima, J. Phys. Soc. Japan \textbf{56}, 2697 (1987).

\bibitem{Khomyakov_2009}
P.A. Khomyakov, G. Giovannetti, P.C. Rusu, G. Brocks, J. van den Brink, and P.J. Kelly, Phys. Rev B \textbf{79}, 195425 (2009)

\bibitem{Farmer_2009}
D. B. Farmer, R. Golizadeh-Mojarad, V. Perebeinos, Y. M. Lin, G. S. Tulevski, J. C. Tsang, and P. Avouris, Nano Lett.
\textbf{9}, 388 (2009).

\end{thebibliography}
\end{document}